\documentclass[aps,prd,preprintnumbers,nofootinbib,preprint]{revtex4}
\usepackage{amsmath,amsfonts,amssymb}
\usepackage{graphicx,epstopdf}
\usepackage{hyperref}

\begin{document}

\title{GUP-Corrected Thermodynamics for all Black Objects and the Existence of Remnants}
 
\author{Mir Faizal}
\email[]{f2mir@uwaterloo.ca}
\affiliation{Department of Physics and Astronomy, University of Waterloo,\\Waterloo, Ontario, N2L 3G1, Canada}

\author{Mohammed M. Khalil}
\email[]{moh.m.khalil@gmail.com}
\affiliation{Department of Electrical Engineering, Alexandria University, Alexandria 12544, Egypt}

\keywords{generalized uncertainty principle; black hole thermodynamics; black hole remnant}

\begin{abstract}
Based on the universality of the entropy-area relation of a black hole, and the fact that the generalized uncertainty principle (GUP) adds a logarithmic correction term to the entropy in accordance with most approaches to quantum gravity, we argue that the GUP-corrected entropy-area relation is universal for all black objects. This correction to the entropy produces corrections to the thermodynamics. We explicitly calculate these corrections for three types of black holes: Reissner--Nordstr\"{o}m, Kerr, and charged AdS black holes, in addition to spinning black rings. In all cases we find that they produce a remnant. Even though the GUP-corrected  entropy-area relation produces the logarithmic term in the series 
expansion, we  need to use the full form of the  GUP-corrected entropy-area relation to 
get remnants for these black holes. 
\end{abstract}

\maketitle

\section{Introduction}

The existence of a minimum measurable length scale of the order of the Planck scale is a universal prediction in almost all approaches to quantum gravity \cite{Amati:1988tn,Garay:1994en,Maggiore:1993rv,Maggiore:1993zu,Maggiore:1993kv,Scardigli:1999jh,Rovelli:1994ge}. For example, in string theory, strings are the smallest probe that can be used for analyzing regions of spacetime, and so it is not possible to probe 
spacetime below the string length scale \cite{Amati:1988tn,Benczik:2002tt}. Hence, string theory comes naturally equipped with a minimum length scale. Also in loop quantum gravity there exists a minimum length scale, which turns the big bang into a big bounce \cite{Dzierzak:2008dy,Ashtekar:2011ni}. 
In fact, there  are strong indications from black hole physics that any theory of quantum gravity should be equipped with a minimum length scale of the order of the Planck length \cite{Maggiore:1993rv, Park:2007az}. This is because the energy required to probe any region of spacetime 
at a smaller scale is more than the energy required to form a black hole in that region.

The existence of a minimum length scale is not consistent with the usual uncertainty principle of quantum mechanics, because according to the usual uncertainty principle, length can be measured to an arbitrary precision if momentum is not measured. It is possible to modify the usual uncertainty principle to a generalized uncertainty principle (GUP), such that this new GUP is consistent with the existence of a minimum length scale  \cite{Garay:1994en,Maggiore:1993rv,Maggiore:1993zu,Maggiore:1993kv,Scardigli:1999jh}. 
Since the uncertainty principle is closely related to the Heisenberg algebra, the GUP deforms the Heisenberg algebra \cite{Kempf:1994su,Kempf:1996fz,Brau:1999uv,nozari2007some,Bambi:2007ty}, which 
in turn changes  the coordinate  representation of the momentum operator \cite{Nozari:2005ex,Nozari:2005mr,Pedram:2010zz,Das:2008kaa}. 
The most studied form of the GUP includes a quadratic term in momentum, and takes the form
\begin{equation}
\label{gup}
\Delta x\Delta p\geq \frac{\hbar}{2}\left(1+\beta \Delta p^2\right)
\end{equation}
where $\beta=\beta_0 l_p^2/\hbar^2$, $\beta_0$ is a dimensionless constant, and $l_p$ is the Planck length.

This generalized uncertainty principle also modifies the thermodynamics of black holes. This modification was calculated for the Schwarzschild black hole in \cite{Adler:2001vs,Medved:2004yu,Myung:2006qr,Nouicer:2007jg,Ali:2012hp}. This is done by first writing the bound on the maximum momentum in terms of energy. This energy is viewed as the energy of an emitted photon, and so, it is related to the black hole temperature. 
Finally, the uncertainty in the position is taken to be proportional to the Schwarzschild radius. This way we get a modified temperature of the Schwarzschild black hole from GUP,  which can in turn be used for calculating the modification to other thermodynamic quantities. 
Thus, we can calculate the corrections to the 
  entropy of a Schwarzschild black hole due to GUP. 

Motivated by the universality of  the leading order correction to the
entropy of a black holes, we   propose that 
the full form of   entropy corrected by GUP is also universal for various black holes. 
So, in this paper, we  argue that the   modified entropy-area relation for the Schwarzschild black hole holds for all black objects.
We apply this relation to Reissner--Nordstr\"{o}m black hole (BH),  Kerr BH, charged AdS BH, and spinning black ring. We calculate their temperature and heat capacity and find they also end up in a remnant. These results are different from those found in Refs. \cite{Xiang:2009yq,Gangopadhyay:2013ofa}, which followed different approaches from the one in this paper. Also, in Ref. \cite{Nozari:2012nf}, the authors used the GUP to predict a remnant for Schwarzschild black hole by studying the tunneling of massless particles.  It may be noted that even though this  modified entropy-area relation produces a logarithmic term in the series 
expansion, we will need to use the full form of the modified entropy-area relation to 
get remnants for various black holes.

\section{Schwarzschild Black Hole}
In this section, we review the derivation of the thermodynamic quantities of the Schwarzschild BH.
We start by solving the GUP in Eq. \eqref{gup} for $\Delta p$
\begin{equation}
\label{deltap}
\Delta p \geq \frac{\Delta x-\sqrt{\Delta x^2-\beta}}{\beta},
\end{equation}
where from now on we use units in which $\hbar=G=c=1$. This expression can be translated to a lower bound on the energy, which can be viewed as the characteristic temperature of a photon emitted from the black hole \cite{Adler:2001vs, Myung:2006qr}
\begin{equation}
T = \frac{\Delta x-\sqrt{\Delta x^2-\beta}}{\beta}.
\end{equation}

The uncertainty in position can be argued to be proportional to the horizon radius $\Delta x=\alpha r_h$ \cite{Adler:2001vs,Medved:2004yu,AmelinoCamelia:2004xx}, where the constant $\alpha=2\pi$ to produce the standard Hawking temperature $T_H=1/4\pi r_h$ when $\beta\to0$.  Thus, we get the modified temperature
\begin{equation}
\label{tempgup}
T= \frac{2\pi r_h-\sqrt{(2\pi r_h)^2-\beta}}{\beta}.
\end{equation}
The temperature becomes complex and unphysical when $r_h<\sqrt{\beta}/2\pi$, which corresponds to the minimum mass 
\begin{equation}
M_{min}=\frac{M_p}{4\pi}\sqrt{\beta},
\end{equation}
where $M_p$ is the Planck mass, and $r_h=2M$. This means the black hole ends in a remnant. 

The entropy can be calculated from the first law of black hole thermodynamics, which for Schwarzschild black holes takes the form
\begin{equation}
S=\int\frac{1}{T}dM=\int\frac{1}{T}\frac{dr_h}{2},
\end{equation}
leading to the modified entropy
\begin{equation}
\label{schent}
S=\frac{2\pi r_h\left(2\pi r_h +\sqrt{4\pi^2 r_h^2-\beta}\right)-\beta\ln\left(2\pi r_h+\sqrt{4\pi^2 r_h^2-\beta}\right)}{8\pi}.
\end{equation}
The heat capacity is calculated from the entropy and temperature via the relation
\begin{equation}
C=T\frac{\partial S}{\partial T}=T\frac{\partial S/\partial r_h}{\partial T/\partial r_h},
\end{equation}
leading to
\begin{equation}
C=\frac{-1}{4\pi}\left(4\pi^2 r_h^2-\beta+2\pi r_h\sqrt{4\pi^2 r_h^2-\beta}\right).
\end{equation}
The heat capacity goes to zero at $r_h=\sqrt{\beta}/2\pi$, which means that the black hole stops exchanging heat with the surrounding space, confirming the existence of a remnant.

\section{The Entropy-Area relation}
We can re-express the entropy relation in Eq. \eqref{schent} in terms of the area of the Schwarzschild black hole $A=4\pi r_h^2$ to get the entropy-area relation
\begin{eqnarray}
\label{modentropy}
S&=&\frac{\pi A+\sqrt{\pi A}\sqrt{\pi A-\beta}-\beta\ln\left(\sqrt{\pi A}+\sqrt{\pi A-\beta}\right)}{8\pi}\nonumber \\ 
&=& S_0 + S_{c}, 
\end{eqnarray}
where $S_0 = A/ 4$ is the original entropy, and $S_c$ are the corrections to this original entropy coming from GUP. 
In the limit in which there is no minimum length, the parameter 
$\beta\to 0$, this expression reduces to the usual expression for entropy, i.e. $S = A/4$. 
Furthermore, the series expansion of this expression to first order is
\begin{equation}
\label{entarea1}
S=\frac{A}{4}-\frac{\beta}{16\pi}\left(\ln(4\pi A)+1\right).
\end{equation}
This in this limit $S_c =  - \beta (\ln(4\pi A)+1)/ 16 \pi  $. 

We argue that the expression for entropy in Eq. \eqref{modentropy} is general for all black objects. In the absence of minimum length, the first term in Eq. \eqref{modentropy}, $S = A/4$, has been verified for all black objects \cite{Altamirano:2014tva}, and any theory of quantum gravity is required to reproduce this term in the black hole entropy. Furthermore, the logarithmic  correction term also seems to occur in most approaches to quantum gravity \cite{Medved:2005vw,Chen:2009sp,Banerjee:2010qc,Sen:2012dw,Bagchi:2013qva,Keeler:2014bra,Nozari:2006vn}, although they differ in the constant factor before the logarithmic term \cite{Ahmad:2012de}. In these theories the logarithmic term seems to be independent of any particular type of black holes. It is an interesting feature that corrections to the entropy generated from the GUP also include terms proportional to the logarithm of the area.

Motivated by the fact that the first term in the expression for the entropy universally has form $S = A/4$,  and the leading order corrections to the entropy are logarithmic corrections, we assume that the generalized uncertainty principle also universally corrects the entropy by a term proportional to the logarithm of the area. Thus, we assume the corrections to the  entropy of all black objects have the form given by Eq. \eqref{modentropy}. The difference for different black objects comes from the exact relation of the area to different quantities, namely mass, charge, and angular momentum. Thus, this correction of entropy in terms of area can be used to calculate the thermodynamic properties of any black hole. 

Furthermore, a difference can occur in the coefficient of this logarithmic term. However, we have   $\beta=\beta_0 l_p^2/\hbar^2$, and there in an arbitrariness in the definition of $\beta_0$. In fact,  there is a large range of allowed values for $\beta_0$ \cite{Das:2008kaa}. So, for each specific case, the difference in the coefficient can be absorbed in the definition of $\beta_0$. The exact value of this coefficient  will not change the existence of a remnant, but only shift the scale at which this remnant is formed. However,  the   interesting conclusion of this analysis  is the existence of a remnant, so, a different value of  coefficient for different black holes, will not change the main conclusion of this paper. 
It may be noted that the existence of the remnant depends critically on the full form of the entropy corrected by GUP. 
So, even though we have argued for arbitrariness in the value of $\beta$ based on the logarithmic corrections to the entropy, 
  we will use the full form of the entropy corrected by GUP. Thus, the main proposal of this paper is that apart from the
arbitrariness in value of the constant $\beta$, the full form 
of the entropy corrected by GUP is  universal, as the leading order corrections terms in its series expansion are universal.

This idea will also hold for exotic black objects like black rings and black Saturn. So, in this paper, we give a proposal for calculating the effect of the generalized uncertainty principle on the thermodynamics of all black objects. We explicitly demonstrate this for three types of black holes: Reissner--Nordstr\"{o}m BH, Kerr BH, and charged AdS BH, in addition to spinning black ring.

\section{Reissner--Nordstr\"{o}m Black Hole}
The Reissner--Nordstr\"{o}m black hole is a spherically symmetric static BH with charge $Q$. The metric takes the form \cite{Altamirano:2014tva}
\begin{equation}
ds^2=-f(r)dt^2+\frac{dr^2}{f(r)}+r^2d\Omega^2,
\end{equation}
where
\begin{equation}
f(r)=1-\frac{2M}{r}+\frac{Q^2}{r^2}.
\end{equation}
The horizon area is given by $A=4\pi r_h^2$, where the horizon radius $r_h$ is the largest solution to $f(r)=0$. The mass can be expressed in terms of $r_h$ and $Q$ via
\begin{equation}
M=\frac{r_h^2-Q^2}{2r_h}.
\end{equation}

The modified temperature of the BH due to the GUP can be calculated from 
\begin{equation}
\label{temp}
\frac{1}{T}=\frac{\partial S}{\partial M}=\frac{\partial S/\partial r_h}{\partial M /\partial r_h}.
\end{equation}
Using the general entropy-area relation from Eq. \eqref{modentropy} we find the temperature
\begin{equation}
T=\frac{(r_h^2-Q^2)\sqrt{4\pi^2r_h^2-\beta}}{r_h^2\left(4\pi^2r_h^2-\beta+2\pi r_h\sqrt{4\pi^2r_h^2-\beta}\right)}.
\end{equation}
This relation goes to zero at $r_h=\sqrt{\beta}/2\pi$ and does not have a physical meaning below this value, signaling the existence of a remnant. This can be further confirmed from the heat capacity.

\begin{figure}[ht]
\centering
\begin{minipage}[b]{0.48\linewidth}
\includegraphics[width=\linewidth]{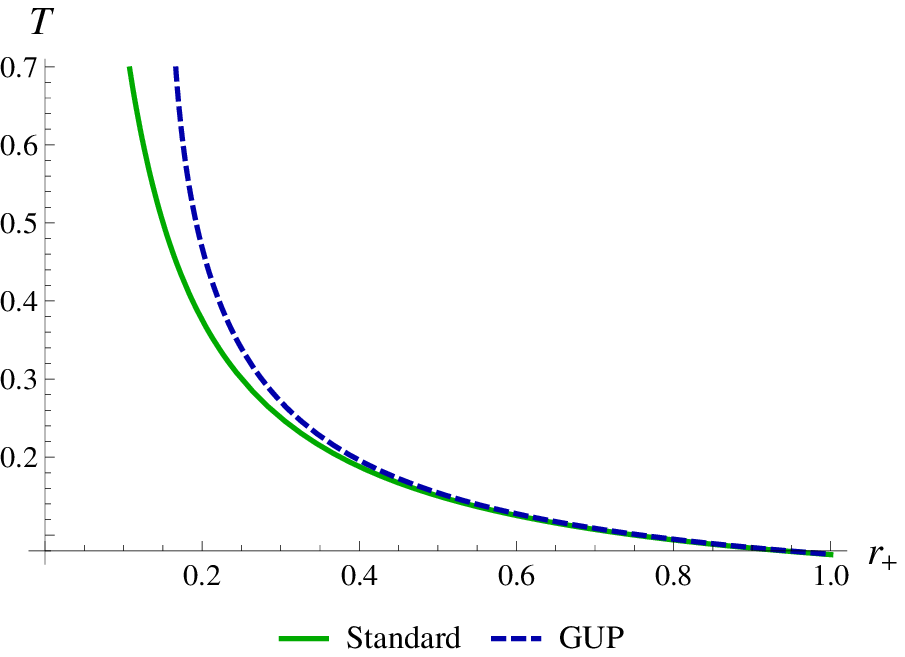}
\caption{\label{fig:RNtemp} Standard and modified temperature of Reissner--Nordstr\"{o}m black hole.}
\end{minipage}
\quad
\begin{minipage}[b]{0.48\linewidth}
\includegraphics[width=\linewidth]{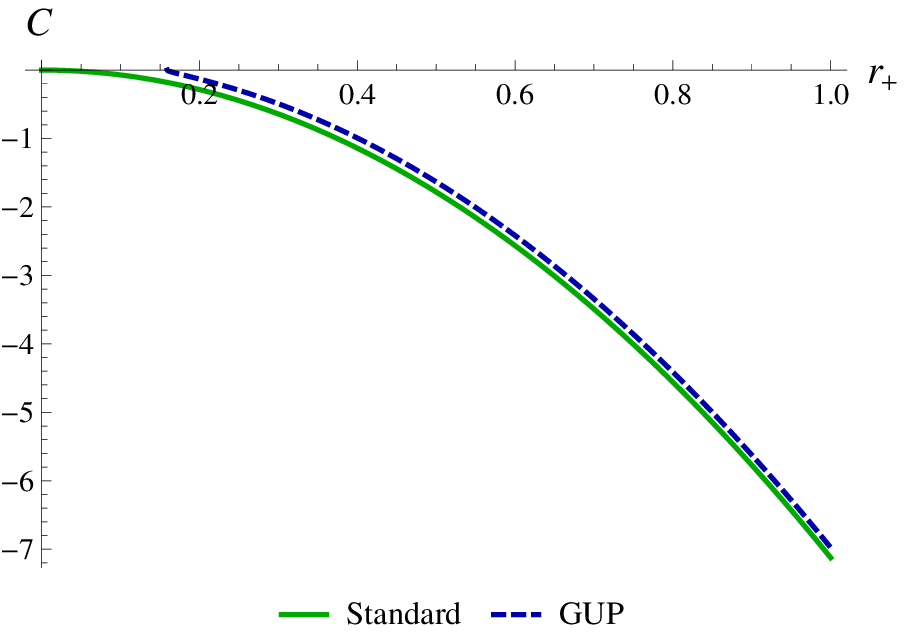}
\caption{\label{fig:RNcap} Standard and modified heat capacity of Reissner--Nordstr\"{o}m black hole.}
\end{minipage}
\end{figure}

The heat capacity is calculated via the relation
\begin{equation}
\label{cap}
C=T\frac{\partial S}{\partial T}=T\frac{\partial S/\partial r_h}{\partial T/\partial r_h}. 
\end{equation}
It may be noted that heat capacity will get corrected as the original entropy $S_0$ has been corrected to   $S= S_0 + S_c$, where $S_c$ are the correction coming from GUP, 
\begin{equation}
C=\frac{-r_h(r_h^2-Q^2)\left(4\pi^2r_h^2-\beta+2\pi r_h\sqrt{4\pi^2r_h^2-\beta}\right)}{4\sqrt{4\pi^2r_h^2-\beta}\left(\pi^2(2r_h^4-6r_h^2Q^2)+\pi r_h(r_h^2-3Q^2)\sqrt{4\pi^2r_h^2-\beta}+Q^2\beta\right)}.
\end{equation}
which goes to zero at $r_h=\sqrt{\beta}/2\pi$ confirming the existence of a remnant, because when the heat capacity is zero, the black hole cannot exchange radiation with the surrounding space. Figures \ref{fig:RNtemp} and \ref{fig:RNcap} are plots of the temperature and heat capacity of the Reissner--Nordstr\"{o}m black hole, and we clearly see the heat capacity goes to zero. In these plots, we assumed $\beta=1$, and that the charge is $Q=M/2$.

\section{Kerr Black Hole}
The Kerr black hole rotates with an angular momentum $J$ characterized by the parameter $a=J/M$. The metric takes the form \cite{Altamirano:2014tva}
\begin{equation}
ds^2=-dt^2+\frac{2Mr}{\Sigma}(dt-a\sin^2\theta d\phi)^2+\frac{\Sigma}{\Delta}dr^2+\Sigma d\theta^2+(r^2+a^2)\sin^2\theta d\phi^2,
\end{equation}
where
\begin{equation}
\Sigma=r^2+a^2\cos^2\theta, \qquad \Delta=r^2+a^2-2Mr.
\end{equation}
The horizon area of the Kerr black hole is given by
\begin{equation}
A=4\pi(r_h^2+a^2),
\end{equation}
where $r_h$ is the horizon radius, and is the largest solution to $\Delta=0$. The mass and angular momentum can be expressed in terms of $r_h$ and $a$ via
\begin{equation}
M=\frac{r_h^2+a^2}{2r_h}, \qquad J=\frac{a}{2r_h}(a^2+r_h^2).
\end{equation}

\begin{figure}[t]
\centering
\begin{minipage}[b]{0.48\linewidth}
\includegraphics[width=\linewidth]{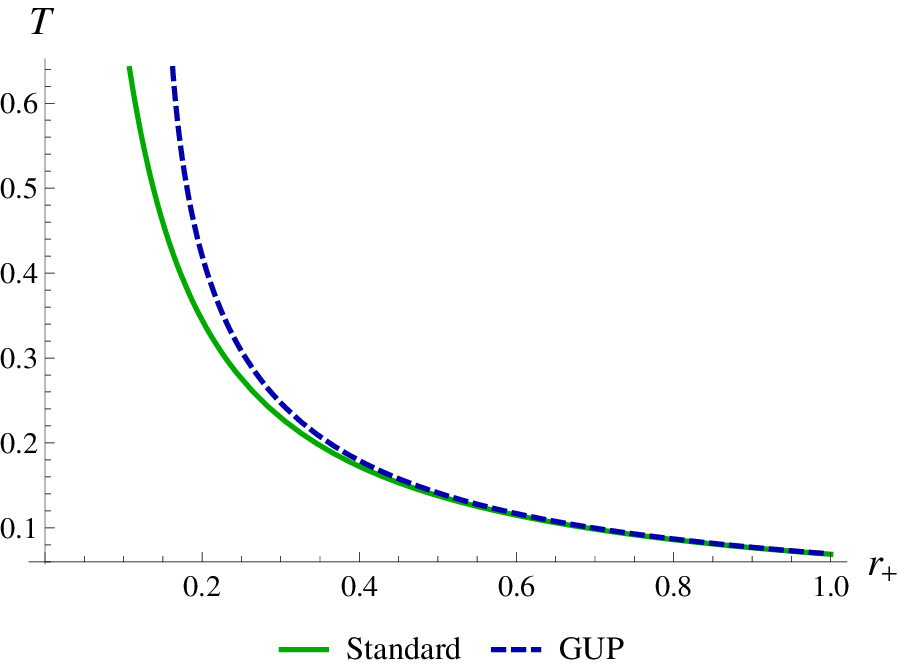}
\caption{\label{fig:kerrtemp} Standard and modified temperature of Kerr black hole.}
\end{minipage}
\quad
\begin{minipage}[b]{0.48\linewidth}
\includegraphics[width=\linewidth]{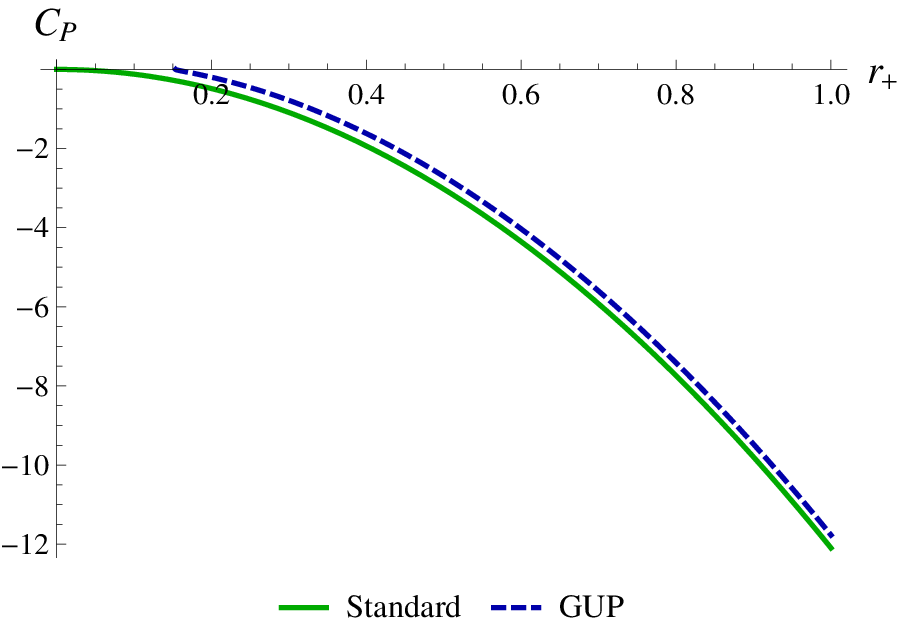}
\caption{\label{fig:kerrcap} Standard and modified heat capacity of Kerr black hole.}
\end{minipage}
\end{figure}

The modified temperature of the BH due to the GUP can be calculated from the entropy-area relation in Eq. \eqref{modentropy} via
\begin{equation}
\label{temprel}
\frac{1}{T}=\left(\frac{\partial S}{\partial M}\right)_J,
\end{equation}
where the partial derivative is calculated at constant angular momentum $J$. This partial derivative can be calculated using the identity
\begin{equation}
\left(\frac{\partial S}{\partial M}\right)_J=\frac{\det\left(\frac{\partial(S,J)}{\partial(r_h,a)}\right)}{\det\left(\frac{\partial(M,J)}{\partial(r_h,a)}\right)},
\end{equation}
where $\frac{\partial(S,J)}{\partial(r_h,a)}$ is the Jacobian matrix, and is defined by
\begin{equation}
\frac{\partial(S,J)}{\partial(r_h,a)}= 
\left[ {\begin{array}{cc}
   \partial S/\partial r_h & \partial S/\partial a \\
   \partial J/\partial r_h & \partial J/\partial a \\
  \end{array} } \right].
\end{equation}
This leads to the modified temperature
\begin{equation}
T=\frac{(r_h^2-a^2)\left(4\pi^2(r_h^2+a^2)-\beta+2\pi\sqrt{r_h^2+a^2}\sqrt{4\pi^2(r_h^2+a^2)-\beta}\right)}{r_h\sqrt{r_h^2+a^2}\sqrt{4\pi^2(r_h^2+a^2)-\beta}\left(8\pi^2(r_h^2+a^2)-\beta+4\pi\sqrt{r_h^2+a^2}\sqrt{4\pi^2(r_h^2+a^2)-\beta}\right)}
\end{equation}
This relation is plotted in figure \ref{fig:kerrtemp}, and it has no physical meaning below $r_h=\sqrt{\beta-4\pi^2a^2}/2\pi$, signaling the existence of a remnant. This is further confirmed by calculating the heat capacity via the relation 
\begin{equation}
C_J=T\left(\frac{\partial S}{\partial T}\right)_J,
\end{equation}
where the derivative is taken at constant $J$ because this is what determines the thermodynamic stability of black holes \cite{Monteiro:2009tc,Altamirano:2014tva}.
\begin{align}
C_J=&\left(\frac{1}{4}\Xi(r_h^4-a^4)\left(8\pi^2a^2+8\pi^2r_h^2+4\pi\Xi-\beta\right)^2\right)/\left[2\pi^2a^6(192\pi^2r^2+24\pi\Xi-13\beta) \nonumber \right. \\ 
&\left.
+\pi r_h^4(6\pi r_h^2\beta+\Xi\beta-16\pi^2 r_h^2\Xi-32\pi^3r_h^4)+a^4(144\pi^3r_h^2\Xi-86\pi^2r_h^2\beta-7\pi\Xi\beta+\beta^2) \nonumber \right.\\
&\left.
+a^2r_h^2(128\pi^4r_h^4+80\pi^3r_h^2\Xi-54\pi^2r_h^2\beta-18\pi\Xi\beta+3\beta^2)+448\pi^4r_h^4a^4+96\pi^4a^8\right]
\end{align}
where, to simplify the expression, we defined
\begin{equation}
\Xi\equiv\sqrt{(r_h^2+a^2)\left(4\pi^2a^2+4\pi^2r_h^2-\beta\right)}.
\end{equation}
The formula for the heat capacity is complicated but what matters is its qualitative behavior, and from figure \ref{fig:kerrcap} we see it goes to zero which confirms the existence of a remnant. In these plots, we assumed $\beta=1$ and $a=M/2$.

\section{Charged A\lowercase{d}S Black Hole}
The metric of charged AdS black holes is given by \cite{Altamirano:2014tva}
\begin{equation}
\label{adsmetric}
ds^2=-f(r)dt^2+\frac{dr^2}{f(r)}+r^2d\theta^2+r^2\sin^2\theta d\phi^2
\end{equation}
with
\begin{equation}
f(r)=1-\frac{2M}{r}+\frac{Q^2}{r^2}+\frac{r^2}{l^2}
\end{equation}
where $Q$ is the charge and $l$ is the AdS radius. The mass $M$ can be expressed in terms of $r_h$ by solving $f(r_+)=0$ leading to
\begin{equation}
M=\frac{r_+^4+l^2(Q^2+r_+^2)}{2r_+l^2}.
\end{equation}

The area of charged AdS black hole is given by $A=4\pi r_h^2$ as in the Schwarzschild case. Thus, the entropy is also given by Eq.\eqref{schent}.
Form the entropy we can calculate the temperature via Eq.\eqref{temp}
\begin{equation}
T=\frac{r_h(3r_h^4+l^2(r_h^2-Q^2))\sqrt{4\pi^2r_h^2-\beta}}{l^2r_h^3\left(4\pi^2r_h^2-\beta+2\pi r_h\sqrt{4\pi^2r_h^2-\beta}\right)}
\end{equation}
Again we notice that the temperature goes to zero at $r_h=\sqrt{\beta}/2\pi$ and does not have a physical meaning below that value, signaling the existence of a remnant. 

The heat capacity is calculated via Eq.\eqref{cap} and leads to
\begin{equation}
C=\frac{r(3r^4+l^2(r^2-Q^2))\left(\xi^2+2\pi r\xi\right)^2} 
{4\xi\left(3r^4\left(2\pi^2r^2-\beta+\pi r\xi\right)-l^2\left(\pi^2(2r^4-6Q^2r^2)+Q^2\beta+\pi r(r^2-3Q^2)\xi\right)\right)},
\end{equation}
where to simplify the above expression we defined $\xi=\sqrt{4\pi^2r^2-\beta}$.
When $r_h=\sqrt{\beta}/2\pi$, the heat capacity also goes to zero which confirms the existence of a remnant. The plots for the temperature and heat capacity are very similar to those of the Kerr black hole.

\section{Spinning Black Ring}
The metric of a spinning neutral black ring in five dimensions takes the form \cite{Emparan:2001wn,Zhao:2006zw,Altamirano:2014tva}
\begin{eqnarray}
\label{ringmetric}
ds^2=&&-\frac{F(y)}{F(x)}\left(dt+C(\nu,\lambda)R\frac{1+y}{F(y)}d\psi\right)^2 \nonumber\\
&& +\frac{R^2F(x)}{(x-y)^2}\left(\frac{dx^2}{G(x)}+\frac{G(x)}{F(x)}d\phi^2-\frac{G(y)}{F(y)}d\psi^2-\frac{dy^2}{G(y)}\right)
\end{eqnarray}
where
\begin{equation}
F(\xi)=1+\lambda\xi,\quad G(\xi)=(1-\xi^2)(1+\nu\xi), \quad
 C(\nu,\lambda)=\sqrt{\lambda(\lambda-\nu)\frac{1+\lambda}{1-\lambda}}.
\end{equation}
The dimensionless parameters $\lambda$ and $\nu$ take values in the range $0<\nu\leq\lambda<1$, and to avoid conical singularity at $x=1$ they must be related by
\begin{equation}
\lambda=\frac{2\nu}{1+\nu^2}.
\end{equation}
The coordinates $x,y$ are restricted to the ranges $-1\leq x\leq 1$ and $-1/\nu\leq y<-1$. The event horizon is located at $y_h=-1/\nu$. The dimensionless parameter $\nu$ determines the shape of the horizon, and can be
considered as a measure for the radius of the ring.

The horizon area is given by \cite{Altamirano:2014tva}
\begin{equation}
A=\frac{8\sqrt{2}\pi^2R^3\nu^2}{(1-\nu)(1+\nu^2)^{3/2}}.
\end{equation}
Thus, the entropy relation \eqref{modentropy} leads to
\begin{equation}
S=\sqrt{2}\pi^2\zeta +2^{-5/4}\sqrt{\pi\zeta}\sqrt{8\sqrt{2}\pi^3\zeta-\beta}- \frac{\beta}{8\pi}\ln\left(2^{7/4}\pi^{3/2}\sqrt{\zeta}- \sqrt{8\sqrt{2}\pi^3\zeta-\beta}\right),
\end{equation}
where to simplify the expression we defined 
\begin{equation}
\zeta\equiv\frac{R^3\nu^2}{(1-\nu)(1+\nu^2)^{3/2}}.
\end{equation}

From the entropy we calculate the temperature using Eq.\eqref{temprel} where $M$ and $J$ are given by
\begin{equation}
\label{mjparameters}
M=\frac{3\pi R^2\nu}{2(1-\nu)(1+\nu^2)},\qquad
J=\frac{\pi\nu R^3}{\sqrt{2}}\left(\frac{1+\nu}{(1-\nu)(1+\nu^2)}\right)^{3/2}.
\end{equation}
This leads to
\begin{equation}
T=\frac{R^2\nu\left(4\pi^2\sqrt{\zeta}-2^{1/4}\sqrt{\pi}\sqrt{8\sqrt{2}\pi^3\zeta-\beta}\right)}{\beta\sqrt{\zeta}(1+\nu^2)}.
\end{equation}
In this case also the temperature goes to zero at a certain value for the parameter $\nu$. This is because $\zeta$ goes to zero when $\nu$ goes to zero. But since the sign of $\beta$ under the square root is negative, the temperature becomes imaginary, which means it reaches zero at a certain value for $\nu$ and stops, i.e. the black ring forms a remnant.  

\section{Conclusions}
It is known that the  relation between the entropy of a black object and the horizon area is always $S = A/4$, 
and the leading  order correction to this entropy is logarithmic. It is also known that the  
generalized uncertainty principle also produces a term proportional to the logarithm of the area. Based on the universality of the relation between the entropy and  the horizon area, we proposed that the expression for the correction to the entropy from the GUP is also universal for all black objects.  
This expression also modifies the thermodynamic quantities of all 
black objects. We explicitly calculated these corrections for three types of black holes: Reissner--Nordstr\"{o}m BH, Kerr BH, and charged AdS BH, in addition to  spinning black ring.  
It was found that in all these cases the black objects end in a remnant. 
Thus, it seems that the GUP predicts a remnant for all black objects. The existence of a remnant can have important phenomenological consequences for the detection of black holes at the LHC \cite{Cavaglia:2003qk,Ali:2012mt,Ali:2014qra}.

The existence of black hole remnants has also been predicted in the context of non-commutative geometry in \cite{Nicolini:2011nz,Mureika:2011hg}, where it is shown that the temperature reaches a maximum value before going to zero at the remnant. Remnants has also been found from gravity's rainbow for all black objects \cite{Ali:2014zea,Ali:2014yea}. Gravity's rainbow is a generalization of doubly special relativity to curved spacetime, and doubly special relativity is based on modifications to the
dispersion relation. Since the GUP also deforms the usual dispersion relation, it seems that the existence of a remnant might be related to the modified dispersion relation. It would be interesting to use the method presented in this paper to investigate the modification of the thermodynamics for other types of black objects. 

\bibliography{gup_kerr}

\begin{thebibliography}{48}
\expandafter\ifx\csname natexlab\endcsname\relax\def\natexlab#1{#1}\fi
\expandafter\ifx\csname bibnamefont\endcsname\relax
  \def\bibnamefont#1{#1}\fi
\expandafter\ifx\csname bibfnamefont\endcsname\relax
  \def\bibfnamefont#1{#1}\fi
\expandafter\ifx\csname citenamefont\endcsname\relax
  \def\citenamefont#1{#1}\fi
\expandafter\ifx\csname url\endcsname\relax
  \def\url#1{\texttt{#1}}\fi
\expandafter\ifx\csname urlprefix\endcsname\relax\def\urlprefix{URL }\fi
\providecommand{\bibinfo}[2]{#2}
\providecommand{\eprint}[2][]{\url{#2}}

\bibitem[{\citenamefont{Amati et~al.}(1989)\citenamefont{Amati, Ciafaloni, and
  Veneziano}}]{Amati:1988tn}
\bibinfo{author}{\bibfnamefont{D.}~\bibnamefont{Amati}},
  \bibinfo{author}{\bibfnamefont{M.}~\bibnamefont{Ciafaloni}},
  \bibnamefont{and}
  \bibinfo{author}{\bibfnamefont{G.}~\bibnamefont{Veneziano}},
  \bibinfo{journal}{Phys. Lett.} \textbf{\bibinfo{volume}{B216}},
  \bibinfo{pages}{41} (\bibinfo{year}{1989}).

\bibitem[{\citenamefont{Garay}(1995)}]{Garay:1994en}
\bibinfo{author}{\bibfnamefont{L.~J.} \bibnamefont{Garay}},
  \bibinfo{journal}{Int.J.Mod.Phys.} \textbf{\bibinfo{volume}{A10}},
  \bibinfo{pages}{145} (\bibinfo{year}{1995}), \eprint{gr-qc/9403008}.

\bibitem[{\citenamefont{Maggiore}(1993{\natexlab{a}})}]{Maggiore:1993rv}
\bibinfo{author}{\bibfnamefont{M.}~\bibnamefont{Maggiore}},
  \bibinfo{journal}{Phys.Lett.} \textbf{\bibinfo{volume}{B304}},
  \bibinfo{pages}{65} (\bibinfo{year}{1993}{\natexlab{a}}),
  \eprint{hep-th/9301067}.

\bibitem[{\citenamefont{Maggiore}(1994)}]{Maggiore:1993zu}
\bibinfo{author}{\bibfnamefont{M.}~\bibnamefont{Maggiore}},
  \bibinfo{journal}{Phys.Rev.} \textbf{\bibinfo{volume}{D49}},
  \bibinfo{pages}{5182} (\bibinfo{year}{1994}), \eprint{hep-th/9305163}.

\bibitem[{\citenamefont{Maggiore}(1993{\natexlab{b}})}]{Maggiore:1993kv}
\bibinfo{author}{\bibfnamefont{M.}~\bibnamefont{Maggiore}},
  \bibinfo{journal}{Phys.Lett.} \textbf{\bibinfo{volume}{B319}},
  \bibinfo{pages}{83} (\bibinfo{year}{1993}{\natexlab{b}}),
  \eprint{hep-th/9309034}.

\bibitem[{\citenamefont{Scardigli}(1999)}]{Scardigli:1999jh}
\bibinfo{author}{\bibfnamefont{F.}~\bibnamefont{Scardigli}},
  \bibinfo{journal}{Phys.Lett.} \textbf{\bibinfo{volume}{B452}},
  \bibinfo{pages}{39} (\bibinfo{year}{1999}), \eprint{hep-th/9904025}.

\bibitem[{\citenamefont{Rovelli and Smolin}(1995)}]{Rovelli:1994ge}
\bibinfo{author}{\bibfnamefont{C.}~\bibnamefont{Rovelli}} \bibnamefont{and}
  \bibinfo{author}{\bibfnamefont{L.}~\bibnamefont{Smolin}},
  \bibinfo{journal}{Nucl.Phys.} \textbf{\bibinfo{volume}{B442}},
  \bibinfo{pages}{593} (\bibinfo{year}{1995}), \eprint{gr-qc/9411005}.

\bibitem[{\citenamefont{Benczik et~al.}(2002)\citenamefont{Benczik, Chang,
  Minic, Okamura, Rayyan et~al.}}]{Benczik:2002tt}
\bibinfo{author}{\bibfnamefont{S.}~\bibnamefont{Benczik}},
  \bibinfo{author}{\bibfnamefont{L.~N.} \bibnamefont{Chang}},
  \bibinfo{author}{\bibfnamefont{D.}~\bibnamefont{Minic}},
  \bibinfo{author}{\bibfnamefont{N.}~\bibnamefont{Okamura}},
  \bibinfo{author}{\bibfnamefont{S.}~\bibnamefont{Rayyan}},
  \bibnamefont{et~al.}, \bibinfo{journal}{Phys.Rev.}
  \textbf{\bibinfo{volume}{D66}}, \bibinfo{pages}{026003}
  (\bibinfo{year}{2002}), \eprint{hep-th/0204049}.

\bibitem[{\citenamefont{Dzierzak et~al.}(2010)\citenamefont{Dzierzak,
  Jezierski, Malkiewicz, and Piechocki}}]{Dzierzak:2008dy}
\bibinfo{author}{\bibfnamefont{P.}~\bibnamefont{Dzierzak}},
  \bibinfo{author}{\bibfnamefont{J.}~\bibnamefont{Jezierski}},
  \bibinfo{author}{\bibfnamefont{P.}~\bibnamefont{Malkiewicz}},
  \bibnamefont{and}
  \bibinfo{author}{\bibfnamefont{W.}~\bibnamefont{Piechocki}},
  \bibinfo{journal}{Acta Phys.Polon.} \textbf{\bibinfo{volume}{B41}},
  \bibinfo{pages}{717} (\bibinfo{year}{2010}), \eprint{0810.3172}.

\bibitem[{\citenamefont{Ashtekar and Singh}(2011)}]{Ashtekar:2011ni}
\bibinfo{author}{\bibfnamefont{A.}~\bibnamefont{Ashtekar}} \bibnamefont{and}
  \bibinfo{author}{\bibfnamefont{P.}~\bibnamefont{Singh}},
  \bibinfo{journal}{Class.Quant.Grav.} \textbf{\bibinfo{volume}{28}},
  \bibinfo{pages}{213001} (\bibinfo{year}{2011}), \eprint{1108.0893}.

\bibitem[{\citenamefont{Park}(2008)}]{Park:2007az}
\bibinfo{author}{\bibfnamefont{M.-i.} \bibnamefont{Park}},
  \bibinfo{journal}{Phys.Lett.} \textbf{\bibinfo{volume}{B659}},
  \bibinfo{pages}{698} (\bibinfo{year}{2008}), \eprint{0709.2307}.

\bibitem[{\citenamefont{Kempf et~al.}(1995)\citenamefont{Kempf, Mangano, and
  Mann}}]{Kempf:1994su}
\bibinfo{author}{\bibfnamefont{A.}~\bibnamefont{Kempf}},
  \bibinfo{author}{\bibfnamefont{G.}~\bibnamefont{Mangano}}, \bibnamefont{and}
  \bibinfo{author}{\bibfnamefont{R.~B.} \bibnamefont{Mann}},
  \bibinfo{journal}{Phys.Rev.} \textbf{\bibinfo{volume}{D52}},
  \bibinfo{pages}{1108} (\bibinfo{year}{1995}), \eprint{hep-th/9412167}.

\bibitem[{\citenamefont{Kempf}(1997)}]{Kempf:1996fz}
\bibinfo{author}{\bibfnamefont{A.}~\bibnamefont{Kempf}},
  \bibinfo{journal}{J.Phys.} \textbf{\bibinfo{volume}{A30}},
  \bibinfo{pages}{2093} (\bibinfo{year}{1997}), \eprint{hep-th/9604045}.

\bibitem[{\citenamefont{Brau}(1999)}]{Brau:1999uv}
\bibinfo{author}{\bibfnamefont{F.}~\bibnamefont{Brau}},
  \bibinfo{journal}{J.Phys.} \textbf{\bibinfo{volume}{A32}},
  \bibinfo{pages}{7691} (\bibinfo{year}{1999}), \eprint{quant-ph/9905033}.

\bibitem[{\citenamefont{Nozari and Fazlpour}(2007)}]{nozari2007some}
\bibinfo{author}{\bibfnamefont{K.}~\bibnamefont{Nozari}} \bibnamefont{and}
  \bibinfo{author}{\bibfnamefont{B.}~\bibnamefont{Fazlpour}},
  \bibinfo{journal}{Chaos, Solitons \& Fractals} \textbf{\bibinfo{volume}{34}},
  \bibinfo{pages}{224} (\bibinfo{year}{2007}).

\bibitem[{\citenamefont{Bambi and Urban}(2008)}]{Bambi:2007ty}
\bibinfo{author}{\bibfnamefont{C.}~\bibnamefont{Bambi}} \bibnamefont{and}
  \bibinfo{author}{\bibfnamefont{F.}~\bibnamefont{Urban}},
  \bibinfo{journal}{Class.Quant.Grav.} \textbf{\bibinfo{volume}{25}},
  \bibinfo{pages}{095006} (\bibinfo{year}{2008}), \eprint{0709.1965}.

\bibitem[{\citenamefont{Nozari}(2005)}]{Nozari:2005ex}
\bibinfo{author}{\bibfnamefont{K.}~\bibnamefont{Nozari}},
  \bibinfo{journal}{Phys.Lett.} \textbf{\bibinfo{volume}{B629}},
  \bibinfo{pages}{41} (\bibinfo{year}{2005}), \eprint{hep-th/0508078}.

\bibitem[{\citenamefont{Nozari and Azizi}(2006)}]{Nozari:2005mr}
\bibinfo{author}{\bibfnamefont{K.}~\bibnamefont{Nozari}} \bibnamefont{and}
  \bibinfo{author}{\bibfnamefont{T.}~\bibnamefont{Azizi}},
  \bibinfo{journal}{Gen.Rel.Grav.} \textbf{\bibinfo{volume}{38}},
  \bibinfo{pages}{735} (\bibinfo{year}{2006}), \eprint{quant-ph/0507018}.

\bibitem[{\citenamefont{Pedram}(2010)}]{Pedram:2010zz}
\bibinfo{author}{\bibfnamefont{P.}~\bibnamefont{Pedram}},
  \bibinfo{journal}{Int.J.Mod.Phys.} \textbf{\bibinfo{volume}{D19}},
  \bibinfo{pages}{2003} (\bibinfo{year}{2010}), \eprint{1103.3805}.

\bibitem[{\citenamefont{Das and Vagenas}(2008)}]{Das:2008kaa}
\bibinfo{author}{\bibfnamefont{S.}~\bibnamefont{Das}} \bibnamefont{and}
  \bibinfo{author}{\bibfnamefont{E.~C.} \bibnamefont{Vagenas}},
  \bibinfo{journal}{Phys.Rev.Lett.} \textbf{\bibinfo{volume}{101}},
  \bibinfo{pages}{221301} (\bibinfo{year}{2008}), \eprint{0810.5333}.

\bibitem[{\citenamefont{Adler et~al.}(2001)\citenamefont{Adler, Chen, and
  Santiago}}]{Adler:2001vs}
\bibinfo{author}{\bibfnamefont{R.~J.} \bibnamefont{Adler}},
  \bibinfo{author}{\bibfnamefont{P.}~\bibnamefont{Chen}}, \bibnamefont{and}
  \bibinfo{author}{\bibfnamefont{D.~I.} \bibnamefont{Santiago}},
  \bibinfo{journal}{Gen.Rel.Grav.} \textbf{\bibinfo{volume}{33}},
  \bibinfo{pages}{2101} (\bibinfo{year}{2001}), \eprint{gr-qc/0106080}.

\bibitem[{\citenamefont{Medved and Vagenas}(2004)}]{Medved:2004yu}
\bibinfo{author}{\bibfnamefont{A.}~\bibnamefont{Medved}} \bibnamefont{and}
  \bibinfo{author}{\bibfnamefont{E.~C.} \bibnamefont{Vagenas}},
  \bibinfo{journal}{Phys.Rev.} \textbf{\bibinfo{volume}{D70}},
  \bibinfo{pages}{124021} (\bibinfo{year}{2004}), \eprint{hep-th/0411022}.

\bibitem[{\citenamefont{Myung et~al.}(2007)\citenamefont{Myung, Kim, and
  Park}}]{Myung:2006qr}
\bibinfo{author}{\bibfnamefont{Y.~S.} \bibnamefont{Myung}},
  \bibinfo{author}{\bibfnamefont{Y.-W.} \bibnamefont{Kim}}, \bibnamefont{and}
  \bibinfo{author}{\bibfnamefont{Y.-J.} \bibnamefont{Park}},
  \bibinfo{journal}{Phys.Lett.} \textbf{\bibinfo{volume}{B645}},
  \bibinfo{pages}{393} (\bibinfo{year}{2007}), \eprint{gr-qc/0609031}.

\bibitem[{\citenamefont{Nouicer}(2007)}]{Nouicer:2007jg}
\bibinfo{author}{\bibfnamefont{K.}~\bibnamefont{Nouicer}},
  \bibinfo{journal}{Phys.Lett.} \textbf{\bibinfo{volume}{B646}},
  \bibinfo{pages}{63} (\bibinfo{year}{2007}), \eprint{0704.1261}.

\bibitem[{\citenamefont{Ali et~al.}(2012)\citenamefont{Ali, Nafie, and
  Shalaby}}]{Ali:2012hp}
\bibinfo{author}{\bibfnamefont{A.~F.} \bibnamefont{Ali}},
  \bibinfo{author}{\bibfnamefont{H.}~\bibnamefont{Nafie}}, \bibnamefont{and}
  \bibinfo{author}{\bibfnamefont{M.}~\bibnamefont{Shalaby}},
  \bibinfo{journal}{Europhys.Lett.} \textbf{\bibinfo{volume}{100}},
  \bibinfo{pages}{20004} (\bibinfo{year}{2012}).

\bibitem[{\citenamefont{Xiang and Wen}(2009)}]{Xiang:2009yq}
\bibinfo{author}{\bibfnamefont{L.}~\bibnamefont{Xiang}} \bibnamefont{and}
  \bibinfo{author}{\bibfnamefont{X.}~\bibnamefont{Wen}},
  \bibinfo{journal}{JHEP} \textbf{\bibinfo{volume}{0910}}, \bibinfo{pages}{046}
  (\bibinfo{year}{2009}), \eprint{0901.0603}.

\bibitem[{\citenamefont{Gangopadhyay et~al.}(2014)\citenamefont{Gangopadhyay,
  Dutta, and Saha}}]{Gangopadhyay:2013ofa}
\bibinfo{author}{\bibfnamefont{S.}~\bibnamefont{Gangopadhyay}},
  \bibinfo{author}{\bibfnamefont{A.}~\bibnamefont{Dutta}}, \bibnamefont{and}
  \bibinfo{author}{\bibfnamefont{A.}~\bibnamefont{Saha}},
  \bibinfo{journal}{Gen.Rel.Grav.} \textbf{\bibinfo{volume}{46}},
  \bibinfo{pages}{1661} (\bibinfo{year}{2014}), \eprint{1307.7045}.

\bibitem[{\citenamefont{Nozari and Saghafi}(2012)}]{Nozari:2012nf}
\bibinfo{author}{\bibfnamefont{K.}~\bibnamefont{Nozari}} \bibnamefont{and}
  \bibinfo{author}{\bibfnamefont{S.}~\bibnamefont{Saghafi}},
  \bibinfo{journal}{JHEP} \textbf{\bibinfo{volume}{1211}}, \bibinfo{pages}{005}
  (\bibinfo{year}{2012}), \eprint{1206.5621}.

\bibitem[{\citenamefont{Amelino-Camelia
  et~al.}(2004)\citenamefont{Amelino-Camelia, Arzano, and
  Procaccini}}]{AmelinoCamelia:2004xx}
\bibinfo{author}{\bibfnamefont{G.}~\bibnamefont{Amelino-Camelia}},
  \bibinfo{author}{\bibfnamefont{M.}~\bibnamefont{Arzano}}, \bibnamefont{and}
  \bibinfo{author}{\bibfnamefont{A.}~\bibnamefont{Procaccini}},
  \bibinfo{journal}{Phys.Rev.} \textbf{\bibinfo{volume}{D70}},
  \bibinfo{pages}{107501} (\bibinfo{year}{2004}), \eprint{gr-qc/0405084}.

\bibitem[{\citenamefont{Altamirano et~al.}(2014)\citenamefont{Altamirano,
  Kubiznak, Mann, and Sherkatghanad}}]{Altamirano:2014tva}
\bibinfo{author}{\bibfnamefont{N.}~\bibnamefont{Altamirano}},
  \bibinfo{author}{\bibfnamefont{D.}~\bibnamefont{Kubiznak}},
  \bibinfo{author}{\bibfnamefont{R.~B.} \bibnamefont{Mann}}, \bibnamefont{and}
  \bibinfo{author}{\bibfnamefont{Z.}~\bibnamefont{Sherkatghanad}},
  \bibinfo{journal}{Galaxies} \textbf{\bibinfo{volume}{2}}, \bibinfo{pages}{89}
  (\bibinfo{year}{2014}), \eprint{1401.2586}.

\bibitem[{\citenamefont{Medved and Vagenas}(2005)}]{Medved:2005vw}
\bibinfo{author}{\bibfnamefont{A.}~\bibnamefont{Medved}} \bibnamefont{and}
  \bibinfo{author}{\bibfnamefont{E.~C.} \bibnamefont{Vagenas}},
  \bibinfo{journal}{Mod.Phys.Lett.} \textbf{\bibinfo{volume}{A20}},
  \bibinfo{pages}{1723} (\bibinfo{year}{2005}), \eprint{gr-qc/0505015}.

\bibitem[{\citenamefont{Chen et~al.}(2011)\citenamefont{Chen, Li, and
  Shao}}]{Chen:2009sp}
\bibinfo{author}{\bibfnamefont{Y.-X.} \bibnamefont{Chen}},
  \bibinfo{author}{\bibfnamefont{J.-L.} \bibnamefont{Li}}, \bibnamefont{and}
  \bibinfo{author}{\bibfnamefont{K.-N.} \bibnamefont{Shao}},
  \bibinfo{journal}{Europhys.Lett.} \textbf{\bibinfo{volume}{95}},
  \bibinfo{pages}{10008} (\bibinfo{year}{2011}), \eprint{0910.5540}.

\bibitem[{\citenamefont{Banerjee et~al.}(2011)\citenamefont{Banerjee, Gupta,
  and Sen}}]{Banerjee:2010qc}
\bibinfo{author}{\bibfnamefont{S.}~\bibnamefont{Banerjee}},
  \bibinfo{author}{\bibfnamefont{R.~K.} \bibnamefont{Gupta}}, \bibnamefont{and}
  \bibinfo{author}{\bibfnamefont{A.}~\bibnamefont{Sen}},
  \bibinfo{journal}{JHEP} \textbf{\bibinfo{volume}{1103}}, \bibinfo{pages}{147}
  (\bibinfo{year}{2011}), \eprint{1005.3044}.

\bibitem[{\citenamefont{Sen}(2013)}]{Sen:2012dw}
\bibinfo{author}{\bibfnamefont{A.}~\bibnamefont{Sen}}, \bibinfo{journal}{JHEP}
  \textbf{\bibinfo{volume}{1304}}, \bibinfo{pages}{156} (\bibinfo{year}{2013}),
  \eprint{1205.0971}.

\bibitem[{\citenamefont{Bagchi and Basu}(2014)}]{Bagchi:2013qva}
\bibinfo{author}{\bibfnamefont{A.}~\bibnamefont{Bagchi}} \bibnamefont{and}
  \bibinfo{author}{\bibfnamefont{R.}~\bibnamefont{Basu}},
  \bibinfo{journal}{JHEP} \textbf{\bibinfo{volume}{1403}}, \bibinfo{pages}{020}
  (\bibinfo{year}{2014}), \eprint{1312.5748}.

\bibitem[{\citenamefont{Keeler et~al.}(2014)\citenamefont{Keeler, Larsen, and
  Lisbao}}]{Keeler:2014bra}
\bibinfo{author}{\bibfnamefont{C.}~\bibnamefont{Keeler}},
  \bibinfo{author}{\bibfnamefont{F.}~\bibnamefont{Larsen}}, \bibnamefont{and}
  \bibinfo{author}{\bibfnamefont{P.}~\bibnamefont{Lisbao}},
  \bibinfo{journal}{Phys.Rev.} \textbf{\bibinfo{volume}{D90}},
  \bibinfo{pages}{043011} (\bibinfo{year}{2014}), \eprint{1404.1379}.

\bibitem[{\citenamefont{Nozari and Sefiedgar}(2007)}]{Nozari:2006vn}
\bibinfo{author}{\bibfnamefont{K.}~\bibnamefont{Nozari}} \bibnamefont{and}
  \bibinfo{author}{\bibfnamefont{A.}~\bibnamefont{Sefiedgar}},
  \bibinfo{journal}{Gen.Rel.Grav.} \textbf{\bibinfo{volume}{39}},
  \bibinfo{pages}{501} (\bibinfo{year}{2007}), \eprint{gr-qc/0606046}.

\bibitem[{\citenamefont{Ahmad and Alam}(2012)}]{Ahmad:2012de}
\bibinfo{author}{\bibfnamefont{S.}~\bibnamefont{Ahmad}} \bibnamefont{and}
  \bibinfo{author}{\bibfnamefont{S.}~\bibnamefont{Alam}}
  (\bibinfo{year}{2012}), \eprint{1207.1317}.

\bibitem[{\citenamefont{Monteiro et~al.}(2009)\citenamefont{Monteiro, Perry,
  and Santos}}]{Monteiro:2009tc}
\bibinfo{author}{\bibfnamefont{R.}~\bibnamefont{Monteiro}},
  \bibinfo{author}{\bibfnamefont{M.~J.} \bibnamefont{Perry}}, \bibnamefont{and}
  \bibinfo{author}{\bibfnamefont{J.~E.} \bibnamefont{Santos}},
  \bibinfo{journal}{Phys.Rev.} \textbf{\bibinfo{volume}{D80}},
  \bibinfo{pages}{024041} (\bibinfo{year}{2009}), \eprint{0903.3256}.

\bibitem[{\citenamefont{Emparan and Reall}(2002)}]{Emparan:2001wn}
\bibinfo{author}{\bibfnamefont{R.}~\bibnamefont{Emparan}} \bibnamefont{and}
  \bibinfo{author}{\bibfnamefont{H.~S.} \bibnamefont{Reall}},
  \bibinfo{journal}{Phys.Rev.Lett.} \textbf{\bibinfo{volume}{88}},
  \bibinfo{pages}{101101} (\bibinfo{year}{2002}), \eprint{hep-th/0110260}.

\bibitem[{\citenamefont{Zhao}(2007)}]{Zhao:2006zw}
\bibinfo{author}{\bibfnamefont{L.}~\bibnamefont{Zhao}},
  \bibinfo{journal}{Commun.Theor.Phys.} \textbf{\bibinfo{volume}{47}},
  \bibinfo{pages}{835} (\bibinfo{year}{2007}), \eprint{hep-th/0602065}.

\bibitem[{\citenamefont{Cavaglia et~al.}(2003)\citenamefont{Cavaglia, Das, and
  Maartens}}]{Cavaglia:2003qk}
\bibinfo{author}{\bibfnamefont{M.}~\bibnamefont{Cavaglia}},
  \bibinfo{author}{\bibfnamefont{S.}~\bibnamefont{Das}}, \bibnamefont{and}
  \bibinfo{author}{\bibfnamefont{R.}~\bibnamefont{Maartens}},
  \bibinfo{journal}{Class.Quant.Grav.} \textbf{\bibinfo{volume}{20}},
  \bibinfo{pages}{L205} (\bibinfo{year}{2003}), \eprint{hep-ph/0305223}.

\bibitem[{\citenamefont{Ali}(2012)}]{Ali:2012mt}
\bibinfo{author}{\bibfnamefont{A.~F.} \bibnamefont{Ali}},
  \bibinfo{journal}{JHEP} \textbf{\bibinfo{volume}{1209}}, \bibinfo{pages}{067}
  (\bibinfo{year}{2012}), \eprint{1208.6584}.

\bibitem[{\citenamefont{Ali et~al.}(2014{\natexlab{a}})\citenamefont{Ali,
  Faizal, and Khalil}}]{Ali:2014qra}
\bibinfo{author}{\bibfnamefont{A.~F.} \bibnamefont{Ali}},
  \bibinfo{author}{\bibfnamefont{M.}~\bibnamefont{Faizal}}, \bibnamefont{and}
  \bibinfo{author}{\bibfnamefont{M.~M.} \bibnamefont{Khalil}}
  (\bibinfo{year}{2014}{\natexlab{a}}), \eprint{1410.4765}.

\bibitem[{\citenamefont{Nicolini and Winstanley}(2011)}]{Nicolini:2011nz}
\bibinfo{author}{\bibfnamefont{P.}~\bibnamefont{Nicolini}} \bibnamefont{and}
  \bibinfo{author}{\bibfnamefont{E.}~\bibnamefont{Winstanley}},
  \bibinfo{journal}{JHEP} \textbf{\bibinfo{volume}{1111}}, \bibinfo{pages}{075}
  (\bibinfo{year}{2011}), \eprint{1108.4419}.

\bibitem[{\citenamefont{Mureika et~al.}(2012)\citenamefont{Mureika, Nicolini,
  and Spallucci}}]{Mureika:2011hg}
\bibinfo{author}{\bibfnamefont{J.}~\bibnamefont{Mureika}},
  \bibinfo{author}{\bibfnamefont{P.}~\bibnamefont{Nicolini}}, \bibnamefont{and}
  \bibinfo{author}{\bibfnamefont{E.}~\bibnamefont{Spallucci}},
  \bibinfo{journal}{Phys.Rev.} \textbf{\bibinfo{volume}{D85}},
  \bibinfo{pages}{106007} (\bibinfo{year}{2012}), \eprint{1111.5830}.

\bibitem[{\citenamefont{Ali et~al.}(2014{\natexlab{b}})\citenamefont{Ali,
  Faizal, and Khalil}}]{Ali:2014zea}
\bibinfo{author}{\bibfnamefont{A.~F.} \bibnamefont{Ali}},
  \bibinfo{author}{\bibfnamefont{M.}~\bibnamefont{Faizal}}, \bibnamefont{and}
  \bibinfo{author}{\bibfnamefont{M.~M.} \bibnamefont{Khalil}}
  (\bibinfo{year}{2014}{\natexlab{b}}), \eprint{1410.5706}.

\bibitem[{\citenamefont{Ali et~al.}(2014{\natexlab{c}})\citenamefont{Ali,
  Faizal, and Khalil}}]{Ali:2014yea}
\bibinfo{author}{\bibfnamefont{A.~F.} \bibnamefont{Ali}},
  \bibinfo{author}{\bibfnamefont{M.}~\bibnamefont{Faizal}}, \bibnamefont{and}
  \bibinfo{author}{\bibfnamefont{M.~M.} \bibnamefont{Khalil}}
  (\bibinfo{year}{2014}{\natexlab{c}}), \eprint{1409.5745}.

\end{thebibliography}

\end{document}